\documentclass[a4paper]{article}

\usepackage{arxiv}

\usepackage[utf8]{inputenc} 
\usepackage[T1]{fontenc}    
\usepackage[final]{graphicx}
\usepackage{hyperref}       
\usepackage{url}            
\usepackage{booktabs}       

\usepackage{microtype}      

\usepackage{calc,ifthen}

\usepackage{amsmath,amsthm,amssymb}
\usepackage{amsfonts}       
\usepackage{nicefrac}       

\usepackage[textsize=normalsize,textwidth=2cm,colorinlistoftodos]{todonotes}

\usepackage{tikz,pgfplots}
\usetikzlibrary{arrows.meta,decorations.pathmorphing,
     decorations.markings,shadows,calc}

\usepackage{cleveref}

\numberwithin{equation}{section}

\providecommand*{\eu}{\ensuremath{\mathrm{e}}}
\providecommand*{\iu}{\ensuremath{\mathrm{i}}}

\renewcommand{\vec}[1]{\boldsymbol{#1}}

\title{An ultrasonic measurement of stress in steel without calibration: the angled shear wave identity}

\author{
  Guo-Yang Li \\
  Harvard Medical School and Wellman Center for Photomedicine\\
  Massachusetts General Hospital\\
  Boston, MA 02114, USA \\
  \texttt{gli26@mgh.harvard.edu} \\
   \And
  Artur L.~Gower\thanks{Webpage: \href{https://arturgower.github.io}{arturgower.github.io}.} \\
  Department of Mechanical Engineering\\
  University of Sheffield\\
  Sheffield, UK \\
  \texttt{arturgower@gmail.com} \\
   \And
 Michel Destrade\\
  School of Mathematics, Statistics and Applied Mathematics\\
   NUI Galway \\
  Galway, Ireland \\
  \texttt{michel.destrade@nuigalway.ie} \\
}

\graphicspath{{images/},{../images/} }

\begin{document}

\maketitle



\begin{abstract}

Measuring stress levels in loaded structures is crucial to assess and monitor their health, and to predict the length of their remaining structural life. However, measuring stress non-destructively has proved quite challenging.
Many ultrasonic methods are able to accurately predict in-plane stresses in a controlled laboratory environment, but struggle to be robust outside, in a real world setting.
That is because they rely either on knowing beforehand the material constants  (which are difficult to acquire) or they require significant calibration for each specimen.
Here we present a simple ultrasonic method to evaluate the in-plane stress \emph{in situ} directly,  without knowing any material constants. This method only requires measuring the speed of two angled shear waves.
It is based on a formula which is exact for incompressible solids, such as soft gels or tissues, and is approximately true for compressible ``hard'' solids, such as steel and other metals.
We validate the formula against virtual experiments using Finite Element simulations, and find it displays excellent accuracy, with a small error of the order of 1\%.

\end{abstract}



\section{Background: The need to monitor stress}


Railroad rails in the real world can degrade greatly because of  high levels of built-up mechanical stresses. During cold winter nights steel rails contract and the resulting tensile stress then promotes fatigue cracks. On hot summer days rails expand and the resulting compressive stress can trigger catastrophic buckling.

\begin{figure}[!ht]
  \centering
  \includegraphics[height=4.5cm]{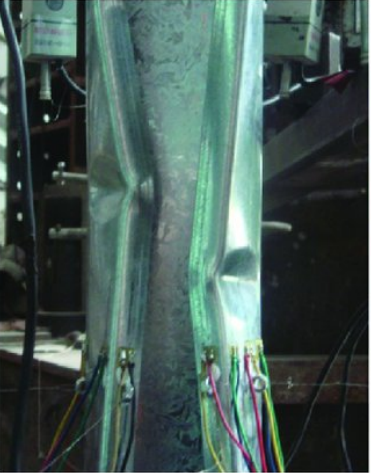}
  \includegraphics[height=4.5cm]{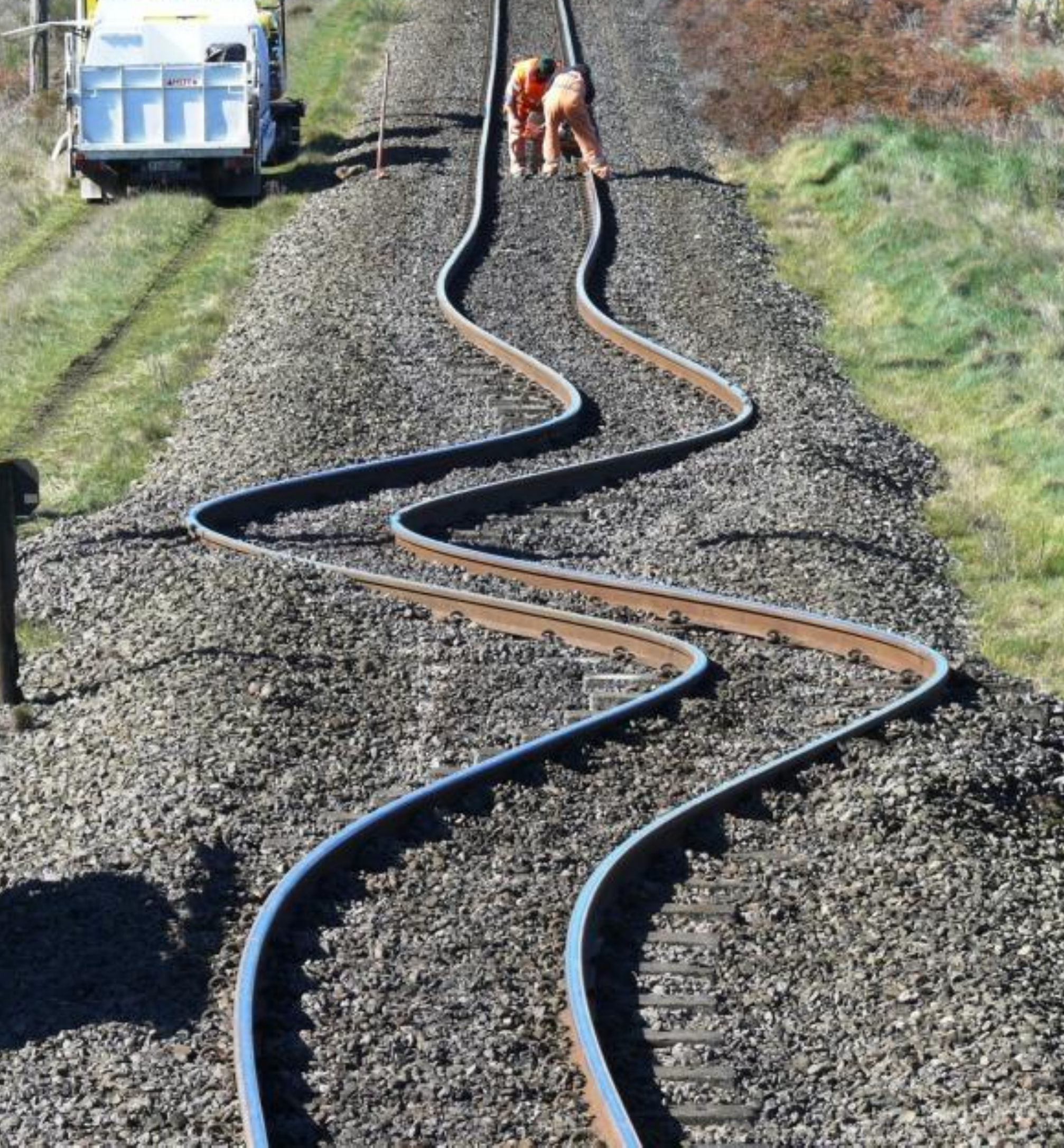}
  \caption{Examples of failure due to compressive stresses. On the left: A steel column which has buckled (in the lab) due to applied compressive stresses ~\cite{li2016plane}. On the right: a rail track which has buckled due to compressive stress following an Earthquake~\cite{buckled-rail}. }
  \label{fig:stress-buckling}
\end{figure}

Rail steel is one of many examples where stress variations and/or high stresses lead to wear and failure \cite{hirao_electromagnetic_2017}.
Other examples include stress in steel pipes~\cite{law2006residual}, pressure vessels, and common machine components~\cite{yakovlev2014measurements} such as bearing raceways. Accurately evaluating the stress within steel plates and bars would substantially improve our estimates on structure life \cite{bray2001subsurface}, and help us to efficiently schedule maintenance and improve safety. \Cref{fig:stress-buckling} shows two examples of steel structures that have failed due to a build up of compressive stress: the left shows a buckled steel column and the right a buckled track. Other than buckling, stress also causes material fatigue and crack growth~\cite{schijve2003fatigue}.

\begin{figure}[!h]
  \centering
  \includegraphics[width=0.55\linewidth]{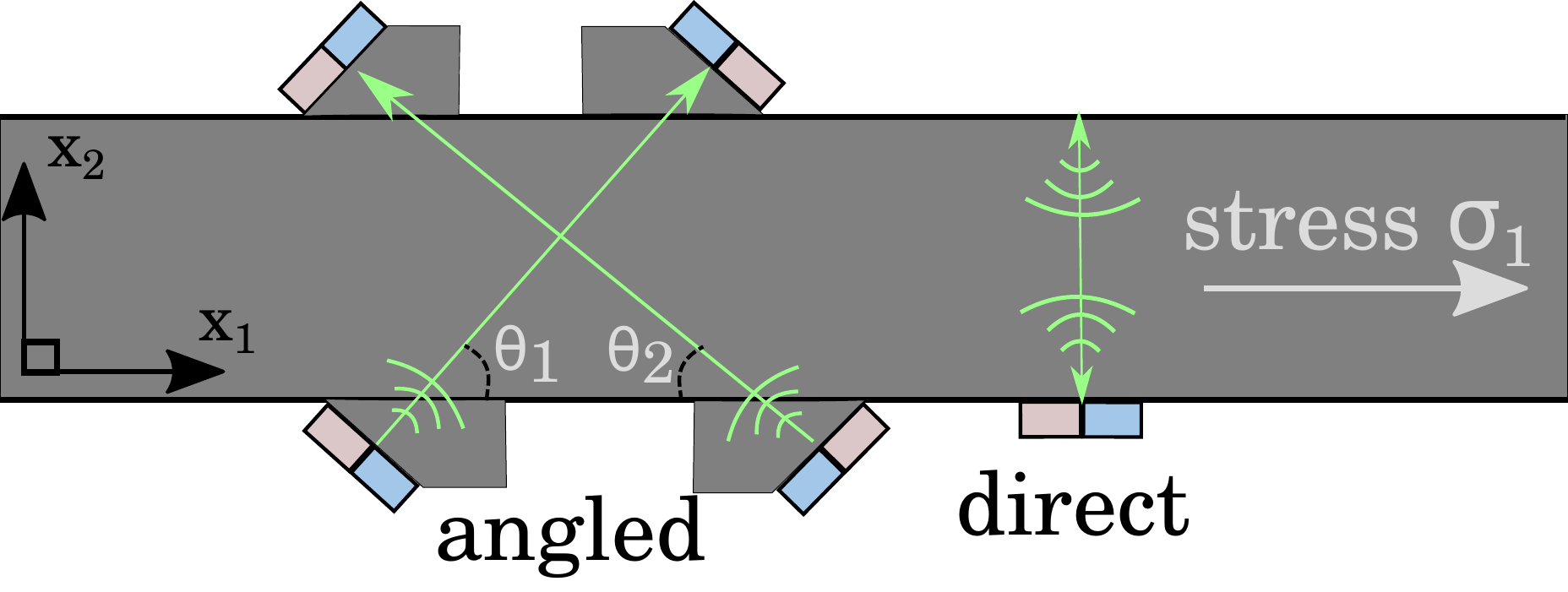}
  \caption{To measure the uni-axial stress $\sigma_1$ we can  send two ultrasonic waves either directly across the steel plate, or at an angle.
  In the former method (birefringence) we need an  \emph{a priori} knowledge of the elastic constants to deduce the stress; by contrast, the angled method (this paper) gives instant access to the stress.}
  \label{fig:web-sensors}
\end{figure}

Stress measurement methods based on ultrasonic elastic wave propagation have long been used as they are relatively cheap, safe, quick and non-destructive~\cite{hirao_electromagnetic_2017,yan2018progress}. The basic setup is shown in \Cref{fig:web-sensors}. In this work we focus on structures made of a hard solid, like steel, where we have access to at least one relatively flat surface, such as the web rail.
The methods to measure stress with ultrasonic waves in other scenarios, such as strands and cables~\cite{chaki2009stress}, can be quite different.

One of the simplest methods to measure stress, in a scenario similar to \Cref{fig:web-sensors}, is called \emph{ultrasonic birefringence}.
It requires measuring the speed of two shear waves travelling directly across the plate.
Then, using the coordinate system shown in \Cref{fig:web-sensors}, assuming only a uni-axial stress along $x_1$, and symmetry along $x_3$, we find that \cite{toupin1961sound,schneider1995ultrasonic, abiza2012large}
\begin{equation} \label{eqn:birefringence-identity}
\rho (v_{21}^2 - v_{23}^2) = \left(1 + \frac{n}{4\mu}\right) \sigma_1, \qquad \text{(Birefringence identity)}
\end{equation}
where $v_{21}$ ($v_{23}$, respectively) is the speed of a shear wave propagating in the $x_2-$direction and polarised in the $x_1-$direction ($x_3-$direction, respectively), and $\rho$ is the current mass density.

\begin{figure}[!ht]
  \centering
  \includegraphics[width=0.6\linewidth]{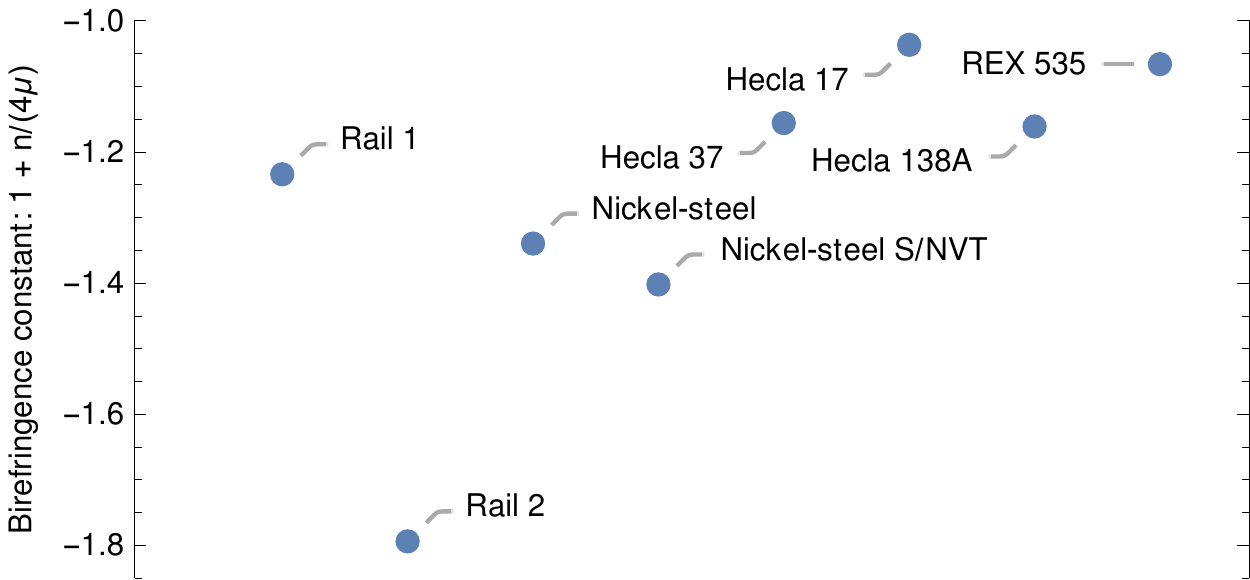}
  \caption{Range of values for the birefringence constant of 10 samples of steel. Note that the constants were all measured in a highly controlled laboratory environments.
  Depending on the type of steel, this constant varies between 10\% (Nickel-steel) and 40\% (Rail steel).
The references and values of the constants are given in~\Cref{tab:material parameters}.}
  \label{fig:birefringence-constants}
\end{figure}

Although these two wave speeds are quite easy to measure, the accuracy of the resulting measurement of $\sigma_1$ is governed by the accuracy of the term in brackets on the right-hand side--called the \emph{birefringence constant}--, which involves $\mu$, a second-order Lam\'e elastic constant, and $n$, a third-order Murnaghan elastic constant.
Birefringence constants are very difficult to measure \emph{in situ}, and instead are measured in the lab, often on pristine steel samples, which obviously can be very different from their in-service and worn counterparts (for instance, these constants change with temperature and wear~\cite{muir2009one}.)
In addition, whereas the mass density $\rho$ and the second-order constants $\lambda$, $\mu$ can be measured quite consistently from one steel sample to another, the measurements of the third-order constants vary widely, as can be checked in \Cref{tab:material parameters}. Even with a perfect measurement of the shear wave speeds, we can expect this variation in the constants to lead to an error of $40\%$, as shown in \Cref{fig:birefringence-constants}, where the birefringence constant of ten samples of steel is reported.

Another notable ultrasonic method which use changes in ultrasonic speeds to measure stress below the material surface is the `longitudinal critically refracted method'~\cite{bray2001subsurface,wang2018improved}. It is mostly used for plates, as its penetration depth is limited to one wavelength. There are further alternative methods for measuring stress along the surface of the material, including using Rayleigh or Lamb waves~\cite{pei2016higher}. These usually require specific surface conditions, and often that the surface of the material is the same as the bulk.
Similar to Birefringence, these methods require prior knowledge of the material constants or significant calibration.
Designing a method to measure stress, without knowing the third order constants has long been a goal worthy of investigation and applications~\cite{thompson1986angular}.


In this paper, we propose a  method to measure stress directly  based on the following approximate identity
\begin{equation} \label{eqn:stress-identity-second-order}
  \rho_0 \frac{v_{1}^2 - v_{2}^2}{\cos(2\theta_1)} \simeq \sigma_1 - \sigma_2 , \quad \text{when} \quad \theta_1 \pm \theta_2 = 90^\circ,  \qquad \text{(Angled shear-wave identity)}
\end{equation}
where $\rho_0$ is the mass density before applying stress, and $v_1$, $v_2$ are the speeds of the quasi-shear waves polarised in the $(x_1, x_2)$ plane and traveling in the directions along $(\cos \theta_1,\sin \theta_1,0)$ and $(\cos \theta_2,\sin \theta_2,0)$, respectively.
\Cref{fig:web-sensors} shows examples of these angled shear waves.

Note that the identity above requires no prior knowledge on the material's elastic constants and, as we shall see, it is accurate for a wide range of stresses in steel, with an error typically around 1\%.





\section{Modelling: the angled shear-wave identity}



  \subsection{Exact formula for incompressible solids}


The problem of a small-amplitude body shear wave travelling in a homogeneously stressed, initially isotropic, \emph{incompressible} solid is quite straightforward to solve.

Consider such a solid, subject to a static uniform Cauchy stress $\vec \sigma$.
Call ($x_1,x_2,x_3$) the axes along the principal directions of stress, $\sigma_1$, $\sigma_2$, $\sigma_3$ the corresponding principal stresses, and $\lambda_1$, $\lambda_2$, $\lambda_3$ the principal stretch ratios, which occur along those directions.
Then consider wave propagation in the ($x_1,x_2$) principal plane, along $\vec n = (\cos \theta, \sin \theta,0)$, say.

Because of incompressibility, only pure shear waves can propagate: one (out-of-plane) pure shear wave polarised along $x_3$, and another (in-plane) pure shear wave polarised along $ (-\sin \theta, \cos \theta,0)$ and travelling with speed $v$ given by \cite{ogden2007incremental, destrade2010third}
\begin{equation}
\rho_0 v^2 = \alpha \cos^4\theta + \delta \cos^2\theta \sin^2\theta + \gamma \sin^4\theta.
\label{c^2}
\end{equation}
Here $\rho_0$ is the (constant) mass density of the solid, $\delta$ is a complicated function of the derivatives of the strain energy density $W$ and the stretch ratios (but it is not needed), and
\begin{equation}
\alpha = \dfrac{\sigma_1-\sigma_2}{\lambda_1^2 - \lambda_2^2}\lambda_1^2,
\qquad
\gamma = \dfrac{\sigma_1-\sigma_2}{\lambda_1^2 - \lambda_2^2}\lambda_2^2.
\end{equation}


Then use Equation \eqref{c^2} to compute in turn the squared waves speeds $v_1^2$ and $v_2^2$ of the waves travelling in the $\theta_1=\theta$ and in the $\theta_2=\pi/2 - \theta$ directions, respectively.
Then, by subtraction, the $\delta$ term disappears and we are left with the following \emph{exact formula},
\begin{equation} \label{eqn:incompress-stress-identity}
\rho_0 (v_1^2-v_2^2) =(\sigma_1-\sigma_2)\cos 2\theta.
\end{equation}
Note that this formula is independent of $W$ and gives direct access to the stress difference $\sigma_1-\sigma_2$ once the speeds and the mass density are measured.
It is valid irrespective of the magnitude of the strains and stresses, can prove useful for precise stress assessment in soft, isotropic, incompressible matter, such as  gels \cite{catheline2003measurement, gallot2010time} or some biological tissues \cite{jiang2015measuring, espindola2017flash}.


  \subsection{Approximate formula for compressible solids}


The exact formula \eqref{eqn:incompress-stress-identity} does not translate exactly to a stressed \emph{compressible} solid--such as steel--for two reasons. First, because the mass density now changes with the stresses. 
The second reason is more subtle, but nonetheless known \cite{norris1983propagation}: the in-plane shear wave is no longer a pure shear wave, but instead is a \emph{quasi-shear wave}.
In fact, there are now two in-plane body waves travelling in the $\vec n-$direction: the quasi-shear wave and a quasi-longitudinal wave.

To find their characteristics, we write down and solve the equations governing small-amplitude motion in a homogeneous solid subject to a uniform stress  \cite{dowaikh1991surface,gower2013counter},
\begin{equation} \label{eqn:motion}
\rho \frac{\partial^2 u_{ j}}{\partial t^2} = \mathcal A_{0ijkl} \frac{\partial^2 u_l}{\partial x_k \partial x_{ i}},
\end{equation}
where $\vec u = \vec u(\vec x,t)$ is the mechanical displacement, and  the $\mathcal A_{0ijkl}$ are the so-called \emph{instantaneous elastic moduli} \cite{ogden1997non}.

Looking for in-plane quasi-shear waves of the form $\vec u = (U_1,U_2,0) \eu^{\iu k(x_1 \cos\theta  + x_2 \sin\theta  - v t)}$ where $k$ is the wavenumber, the equation of motion~\eqref{eqn:motion} becomes
\begin{equation} \label{eqn:eigensystem}
\mathbf M
 \begin{bmatrix} U_1 \\ U_2 \end{bmatrix} = \vec 0
   \quad \text{with} \quad
  \mathbf M = \begin{bmatrix}
   \mathcal A_{01111}\cos\theta^2 + \mathcal A_{02121} \sin\theta^2 - \rho v^2 &  (\mathcal A_{01122} +  \mathcal A_{01221})\cos\theta  \sin\theta  \\
    (\mathcal A_{01122} +  \mathcal A_{01221})\cos\theta  \sin\theta  & \mathcal A_{01212}\cos\theta^2 + \mathcal A_{02222} \sin\theta^2 - \rho v^2
  \end{bmatrix}.
 \end{equation}
 We then find the wave speeds by solving
\begin{equation}\label{eqn:secular}
  \det \mathbf M = 0,
\end{equation}
which is a bi-quadratic in $v^2$.

The instantaneous moduli depend on the stresses $\sigma_i$ and on the elastic constants,
 and can be calculated in a straight-forward (albeit long-winded) manner, see details in~\Cref{app:instantaneous-moduli}.
As steel  can sustain only infinitesimal strains in the elastic regime, we adopt the so-called \emph{third-order elasticity} model for its strain-energy density.

As seen in the appendix, the moduli can be computed explicitly at the  level of approximation afforded by third-order elasticity, as
\begin{align} \label{eqn:A}
 & \mathcal A_{01111} = \lambda +  2 \mu + (2 c + 2d + 1) \sigma_1 + (a +  2b) (\sigma_1 + \sigma_2), \notag
  \\
&  \mathcal A_{02222} = \lambda +  2 \mu + (2c + 2d +1) \sigma_2 + (a+2b) (\sigma_1 + \sigma_2),\notag
  \\
 & \mathcal A_{01221} = \mu + \tfrac{1}{2} (2b + d)(\sigma_1 + \sigma_2), \notag\\
 & \mathcal A_{01122} = \lambda + (a + c) (\sigma_1 + \sigma_2),\notag
  \\
&  \mathcal A_{02121} =  \mathcal A_{01221} + \sigma_2, \notag
\\ & \mathcal A_{01212} =  \mathcal A_{01221} + \sigma_1,
\end{align}
where the non-dimensional coefficients $a$, $b$, $c$, $d$ are defined in the appendix (they are constants of order 1, varying in value from -3.6 to + 0.7 for rail steel, see Table \ref{tab:material parameters-2}). A further justification for using the form~\eqref{eqn:A} is that there is significant evidence~\cite{man_hartigs_1998} that for steel the moduli have a linear dependence on the Cauchy stress, even beyond small levels of stress.

Staying within the same level of approximation, we may now solve~\eqref{eqn:secular} for $v^2$, choose the solution  corresponding to a quasi-shear wave, and then expand the result.
We obtain:
\begin{equation} \label{eqn:c-third-order}
\rho v^2 =  \mu +  (b+d/2)(\sigma_1+\sigma_2) +  \sigma_1 \cos^2\theta + \sigma_2 \sin^2\theta.
\end{equation}

Having found the wave speed, we may return to \eqref{eqn:eigensystem} and determine the direction of polarisation.
Instead of being along the $\pi/2 + \theta$ direction, as it would be for a pure shear wave, it is along the $\pi/2+\theta +\theta^\star$ direction for the quasi-shear wave, where $\theta^\star$ is the offset.
Hence, calling $U_0$ the amplitude of the wave, we have $[U_1,U_2] = U_0[-\sin(\theta+\theta^\star), \cos(\theta+\theta^\star)]$.
Substituting into the first  line of the homogeneous system \eqref{eqn:eigensystem}, we conclude that $\tan(\theta+\theta^\star) = M_{12} / M_{11}$, or
\begin{equation} \label{theta-star}
\tan(\theta+\theta^\star) = \left[1 - \frac{c+d}{\lambda + \mu}(\sigma_1 - \sigma_2) \right]\tan\theta.
\end{equation}

Now we go back to \eqref{eqn:c-third-order}, the equation giving the wave speed.
Calling $v_1$ and $v_2$ the speeds of the waves travelling in the $\theta_1 = \theta$ and $\theta_2 = \pi/2 - \theta$ directions, respectively, and subtracting the corresponding equations \eqref{eqn:c-third-order}, we obtain
\begin{equation} \label{eqn:compress-stress-identity}
\rho (v_1^2 - v_2^2) = (\sigma_1-\sigma_2) \cos(2 \theta).
\end{equation}
Finally, we recall that the current mass density $\rho$, which is usually not known, can be related to  $\rho_0$, the initial mass density (measured before applying stress). From~\eqref{eqn:current-density} in the appendix we see that, within the context of third-order elasticity, \eqref{eqn:compress-stress-identity} is equivalent to \eqref{eqn:stress-identity-second-order}, at the same order of approximation.

Note that using \eqref{eqn:c-third-order} to compute the difference of the squared wave speeds of any two shear waves travelling in different directions would also produce a formula giving direct access to the stress without involving the material constants.


\subsection{Wave direction and polarisation}
\label{sec:direction-polarisation}

Practically, there are a number of ways to generate the angled shear waves travelling in the $\theta_1$ and $\theta_2$ directions.
For example, we could apply a shear force on the boundary of a wedge with inclination $\theta$, as shown in~\Cref{fig:web-sensors}.  However, because they are quasi-shear waves, their direction of propagation is not exactly orthogonal to their polarisation, which is to say it is not exactly orthogonal to the wedge face.

In the previous section we saw that the quasi-shear wave propagating in the $\theta$ direction is polarised  in the $\pi/2+\theta+\theta^\star$ direction.
It follows that the quasi-shear wave polarised  in the $\pi/2+\theta$ direction (launched by a wedge with inclination $\theta$) is propagating in the $\theta - \theta^\star$ direction, where $\theta^\star$ is computed from \eqref{theta-star}.
Calling $v_1$ the speed of the wave in the $\theta_1=\theta+\theta^\star$ direction, and $v_2$ that of the wave in the $\theta_2=\pi/2-\theta_1$ direction, we find again that $\rho_0 (v_1^2-v_2^2)=(\sigma_1-\sigma_2)\cos(2\theta)$, because the correction due to the offset is of higher order and negligible in third-order elasticity.
Hence we conclude that the   perturbation in the propagation direction due to the quasi-shear wave character of   the wave  does not affect the prediction of the stress based on  \eqref{eqn:incompress-stress-identity}.

This result is useful, for example, when using finite element software to conduct a virtual experiment, where it is easier to specify the polarisation than the wave direction.


\subsection{Sensitivity to wedge angle errors}


To use the angled shear-wave identity~\eqref{eqn:stress-identity-second-order} and  measure the stress robustly, we need to determine how sensitive it is to experimental errors, including those coming from misaligned wedges, imprecise wave speeds, varying stress levels, among others.
After investigating a number of scenarios, we found that the largest source of imprecision comes from errors in the direction of propagation, such as those arising in the case of misaligned wedges.

\begin{figure}[!ht]
  \centering
  \includegraphics[width=.5\textwidth]{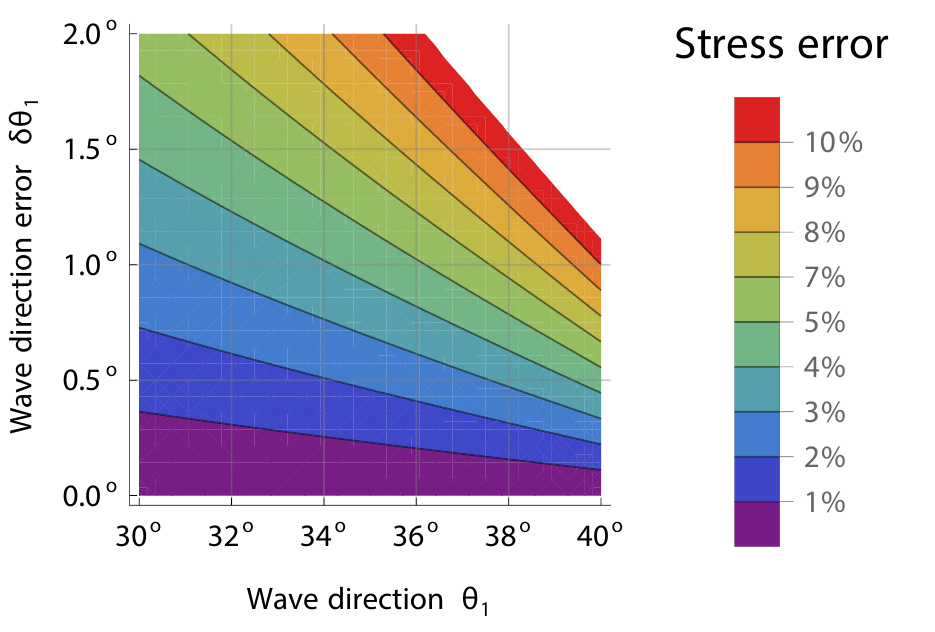}
  \caption{Relative error in the predicted stress $\sigma_1 - \sigma_2$ due to an error in the direction of propagation ($\delta \theta_1$) of one of the shear waves. For example, when one of the waves travels in the direction $\theta_1 = 34^\circ \pm \delta \theta_1$ (while the other wave travels in the direction $\theta_2=56.0^\circ$), we expect: a $4\%$ error in predicting the stress when $\delta \theta_1 \simeq 1^\circ$, and a $7\%$ error in predicting the stress when $\delta \theta_1 \simeq 1.5^\circ$.   }
  \label{fig:angle-errors}
\end{figure}

Assume one of the angled shear waves has a slight error in its direction of propagation, so that instead of travelling in the $\theta_1$ direction it is travelling in the $\theta_1 + \delta \theta_1$ direction.
Call $v_1$ its speed and $v_2$  the  speed of the shear wave propagating at the angle $\theta_2 = \pi/2 - \theta_1$.
Then by perturbation of \eqref{eqn:c-third-order}, we find that the difference between the squared wave speeds is given by
\begin{equation} \label{eqn:angle-error-stress-identity}
  \rho (v_1^2 - v_2^2) = (\sigma_1 - \sigma_2)\left[\cos(2 \theta_1) - \delta \theta_1 \sin(2 \theta_1)\right].
\end{equation}
From this equation we deduce that the relative error in the predicted stress is equal to $\delta \theta_1 \tan(2 \theta_1)$, which we evaluate for various directions $\theta_1$ in \Cref{fig:angle-errors}.
We find that it is small when the precision of the angle $\theta_1$ is of the order of 1$^\circ$ and $\theta_1$ is close to 35$^\circ$.

In the next section, we validate the identity~\eqref{eqn:stress-identity-second-order} further by conducting virtual experiments.



\section{Virtual experiments}

\begin{figure}[!ht]
  \centering
  \includegraphics[width=0.9\linewidth]{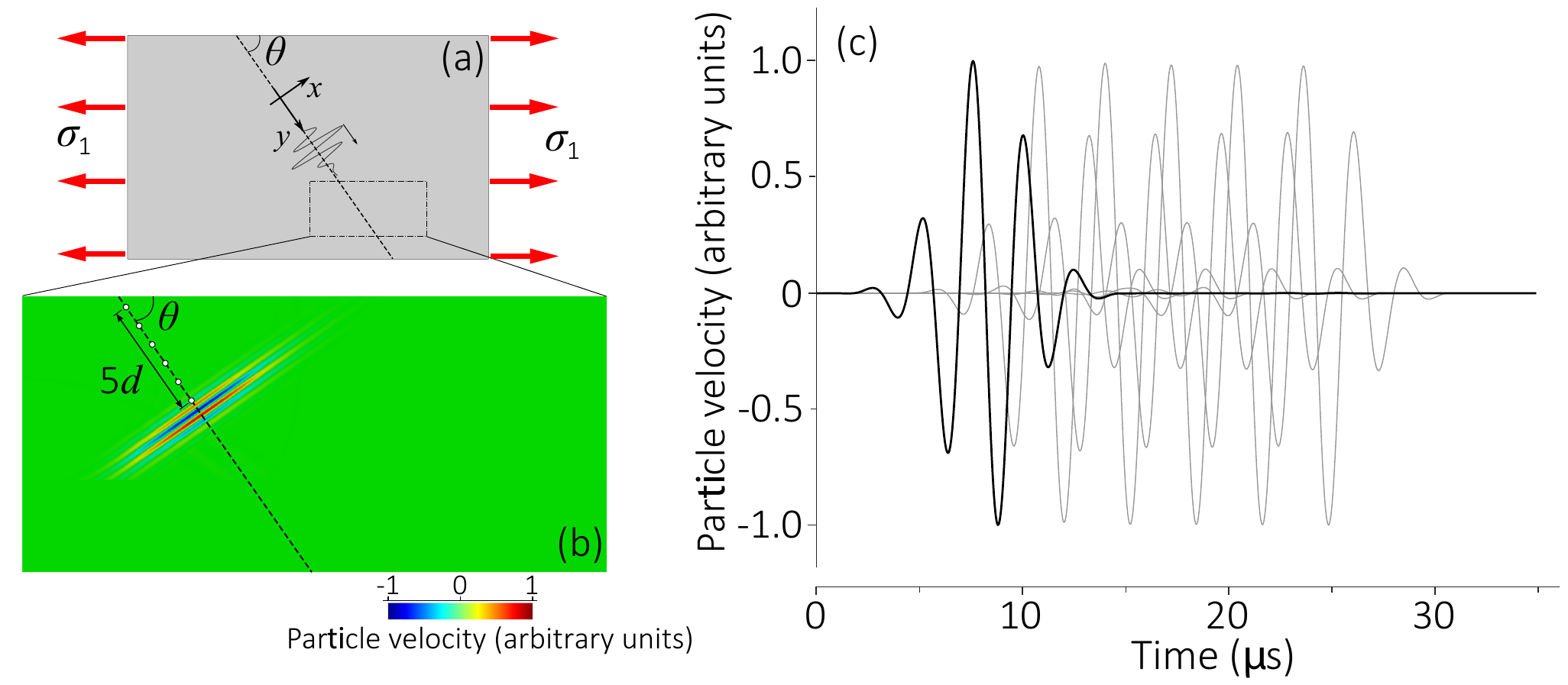}
  \caption{FE simulation of the quasi-shear wave propagation. (a) Schematic of the FE model. The stress is determined by $\sigma_{1}$ and the out-of-plane strain. The wave is induced by a body force defined in the local coordinate systems. (b) Snapshot of the shear wave propagation. Six points at equal distance \emph{d} are used to measure the wave speed. (c) Time profiles of the particle velocities at these points shown in (b).}
  \label{fig:FEA model}
\end{figure}

To validate the method,  we use Finite Element (FE) analysis (Abaqus 6.13, Dassault Systèmes®) to conduct  virtual experiments. We built a plane strain model and created a user-subroutine \emph{UHYPER} to implement a third-order elastic model, such as \eqref{eqn:W-murnaghan}. An illustration of the FE simulation is shown in \Cref{fig:FEA model}(a).

For the material parameters, we choose the values for the sample `Rail steel 1' in \Cref{tab:material parameters}.
We then apply a uni-axial stress along the horizontal direction accompanied by some out-of-plane strain. By prescribing the out-of-plane strain, we can achieve different stress levels and keep the displacement induced by shear wave motion to be zero in this direction. The top and bottom surfaces are stress free.

To describe our setup, let us introduce a local coordinate system $(x,y)$ as shown in \Cref{fig:FEA model}(a), where the $x$ axis is oriented at a $\theta$ angle from the horizontal.
To imitate a real experiment~\cite{schirru2019development}, we generate shear waves by applying an impulse force in the $y$ direction, which would be like applying a shearing force to the face of the wedges shown in \Cref{fig:web-sensors}. We expect this impulse to generate a quasi-shear wave polarised in the $y$ direction, and  propagating in a direction that is not exactly along $x$. We will however assume that the quasi-shear wave does approximately propagate in the $x$ direction. As discussed in \Cref{sec:direction-polarisation}, we expect this assumption to introduce a very small error. In a real experiment, assuming that the quasi-shear wave propagates in the $x$ direction would simplify the experimental setup. So the success of these FE simulations give confidence that this experimental setup would successfully predict the stress.

The impulse we use is a body force $F(x,y;t)$ defined by
\begin{equation}
F(x,y;t) = F_0 \; \mathrm e^{-\left(\dfrac{x^2}{a^2} + \dfrac{y^2}{b^2}\right)} \mathrm e^{-\dfrac{(t-t_0)^2}{\tau^2}}\sin(2\pi ft),
\end{equation}
with $F_0 = 0.1 $ N/mm$^3$, $a = 1$ mm, $b=60$ mm, $t_0 = 6.25$ $\mu$s, $\tau = 2.83$ $\mu$s, and $f = 0.4$ MHz. We point out that we choose to use a rather low frequency $f$ for convenience. We checked that taking higher frequencies leads to the same results, but are much more time-consuming to simulate. The mesh size is approximately $1/15$ of the shear wavelength to ensure the convergence of the simulation.

We use six points spaced with equal distance $d$ along the $x$ axis, as shown in \Cref{fig:FEA model}(b), to measure the shear wave speed.  Here $d$ denotes the distance in the stressed configuration, which is approximately equal to $10$ mm in the initial configuration. In \Cref{fig:FEA model}(c) we plot the time profiles of the particle velocities at these six points. The phase differences between every two adjacent points were measured in the Fourier domain. We then calculate the shear wave speed $v$ from the averaged phase difference $\Delta\phi$ by \cite{li2017mechanics}
\begin{equation}
v = \dfrac{2\pi fd}{\Delta\phi}.
\end{equation}

In the next section we use our FE model to investigate which angle $\theta$ leads the most accurate prediction of the stress, when using our identity~\eqref{eqn:stress-identity-second-order}. We then validate our predictions over a wider range of stress in \Cref{sec:stress-sensitivity}.


\subsection{Sensitivity to wedge angle}


Here we study how the angle of propagation $\theta$ influences the stress predicted by the identity~\eqref{eqn:stress-identity-second-order} when using the wave speeds $v$ simulated by our FE model. For this section, we take the uni-axial stress as $\sigma_{1} = 300$ MPa and no out-of-plane strain: $\lambda_{3} = 1$.

\begin{figure}[!ht]
  \centering
  \includegraphics[width=0.8\linewidth]{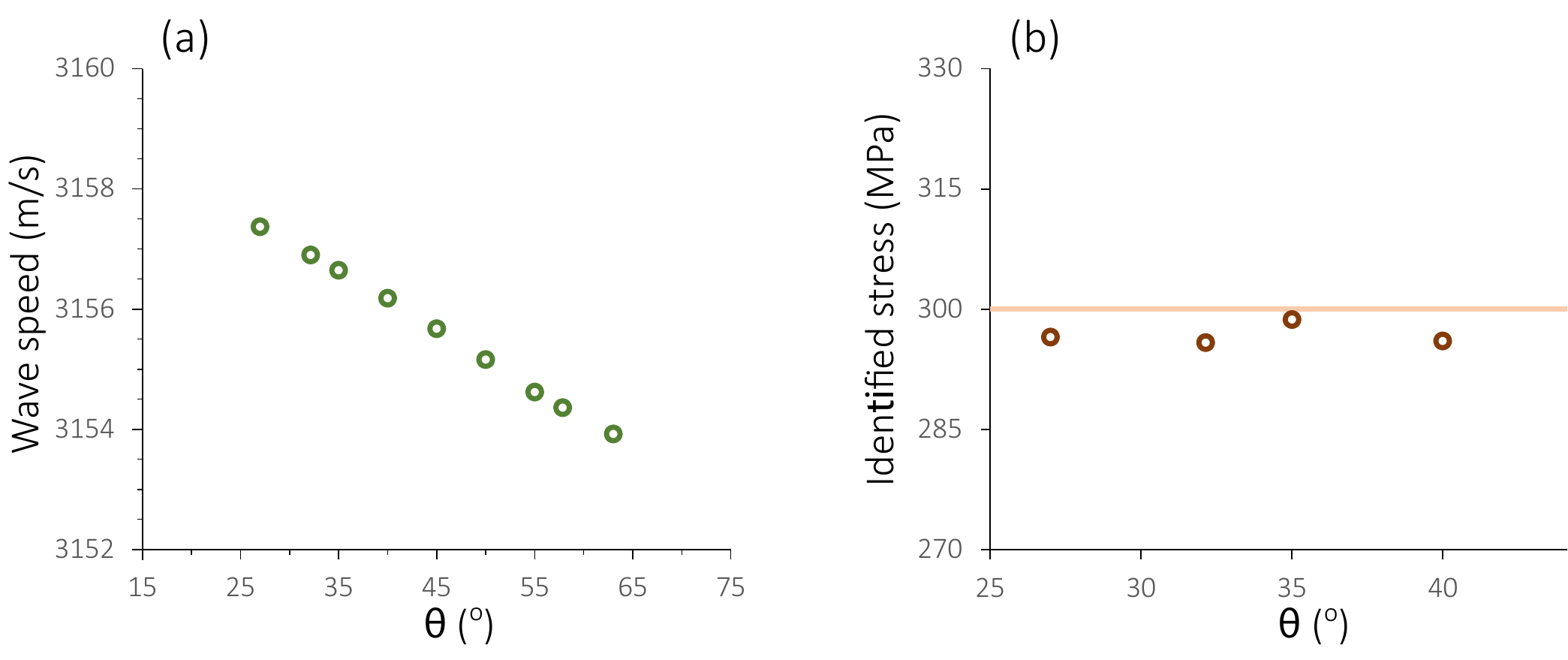}
  \caption{Shear wave speed obtained by the FE simulation and the stress predicited by~\eqref{eqn:stress-identity-second-order}. (a) Shows the how the wave speed depends on the angle $\theta$. (b) Shows the stress predicted by using the FE results fed into~\eqref{eqn:stress-identity-second-order}. The horizontal line denotes the value of $\sigma_{1}$ actually applied to the FE model.}
  \label{fig:Angle dependence}
\end{figure}

As we can see in \Cref{fig:Angle dependence}(a), the shear wave speed $v$ decreases with $\theta$. The wave speeds corresponding to $\theta_1= \theta < 45^\circ$ and $\theta_2 = 90^\circ - \theta$ are $v_1$ and $v_2$ in the identity \eqref{eqn:stress-identity-second-order}, which then yield the stress $\sigma_1$, as shown in \Cref{fig:Angle dependence}(b). For the four different cases $\theta = 27^\circ, 32.14^\circ, 35^\circ, 40^\circ$, the stress identified by \eqref{eqn:stress-identity-second-order} agrees well with the stress $\sigma_{1}$ used in the FE model, with a very small difference ($\leq 1 \%$).

\subsection{Sensitivity to stress levels}
\label{sec:stress-sensitivity}

We now fix the angle shown in~\Cref{fig:FEA model} by choosing $\theta_1=35^\circ, \theta_2=55^\circ$ and study the accuracy of the angled shear wave identity~\eqref{eqn:stress-identity-second-order} for different stress levels. The stress levels we investigate are $\sigma_1 = $100 to 1,500 MPa, which cover all possible stress levels in rail, as standard rail has a maximum tensile strength between 700MPa and 1100MPa~\cite{rail-steel-standards,liberty-rail_2017}.

\begin{figure}[!ht]
  \centering
  \includegraphics[width=0.8\linewidth]{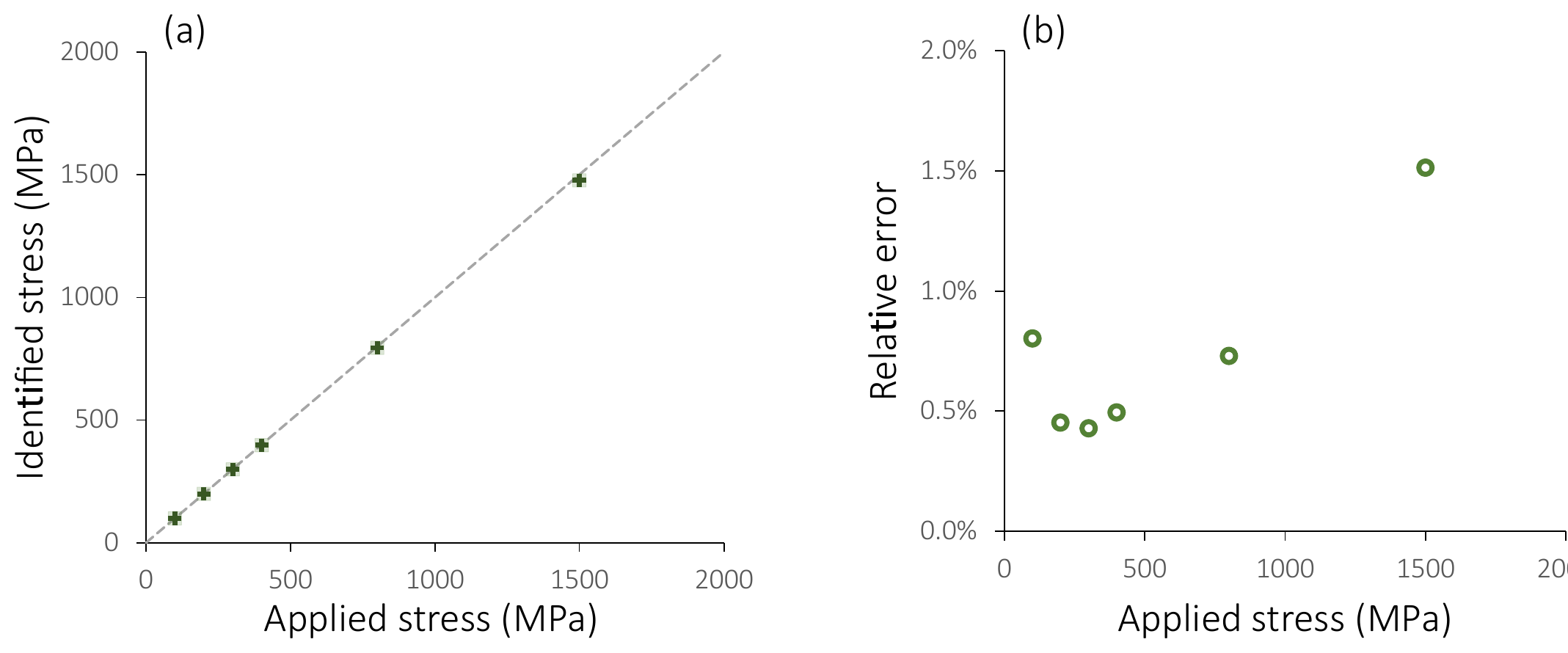}
  \caption{Comparison between the stress applied in the FE simulation and the stress identified by the identity~\eqref{eqn:stress-identity-second-order}. Here the wedges angle, as shown in \Cref{fig:FEA model}, for the two waves are $\theta_1=35^\circ, \theta_2=55^\circ$. (a) Applied stress vs predicted stress. (b) The relative error of the predicted stress. Note that the maximum tensile strength of standard rail steel is maximum tensile strength 1100 MPa.}
  \label{fig:Different stresses}
\end{figure}

As shown in \Cref{fig:Different stresses}, we find that the predicted stresses agree well with the stresses actually applied, with a relative error less than 1.5\%, and an error less than 1\% when below the maximum tensile strength 1100 MPa.

One difference between the FE model and the identity~\eqref{eqn:stress-identity-second-order} is that our FE model imposes an out-of-plane stress $\sigma_3$, where as \eqref{eqn:stress-identity-second-order} has zero out-of-plane stress.
To confirm that this difference has no impact on our predictions we conduct a FE simulation with plane-stress (PS) and then another simulation under plane-strain (PE).

\begin{figure}[!ht]
  \centering
  \includegraphics[width=0.8\linewidth]{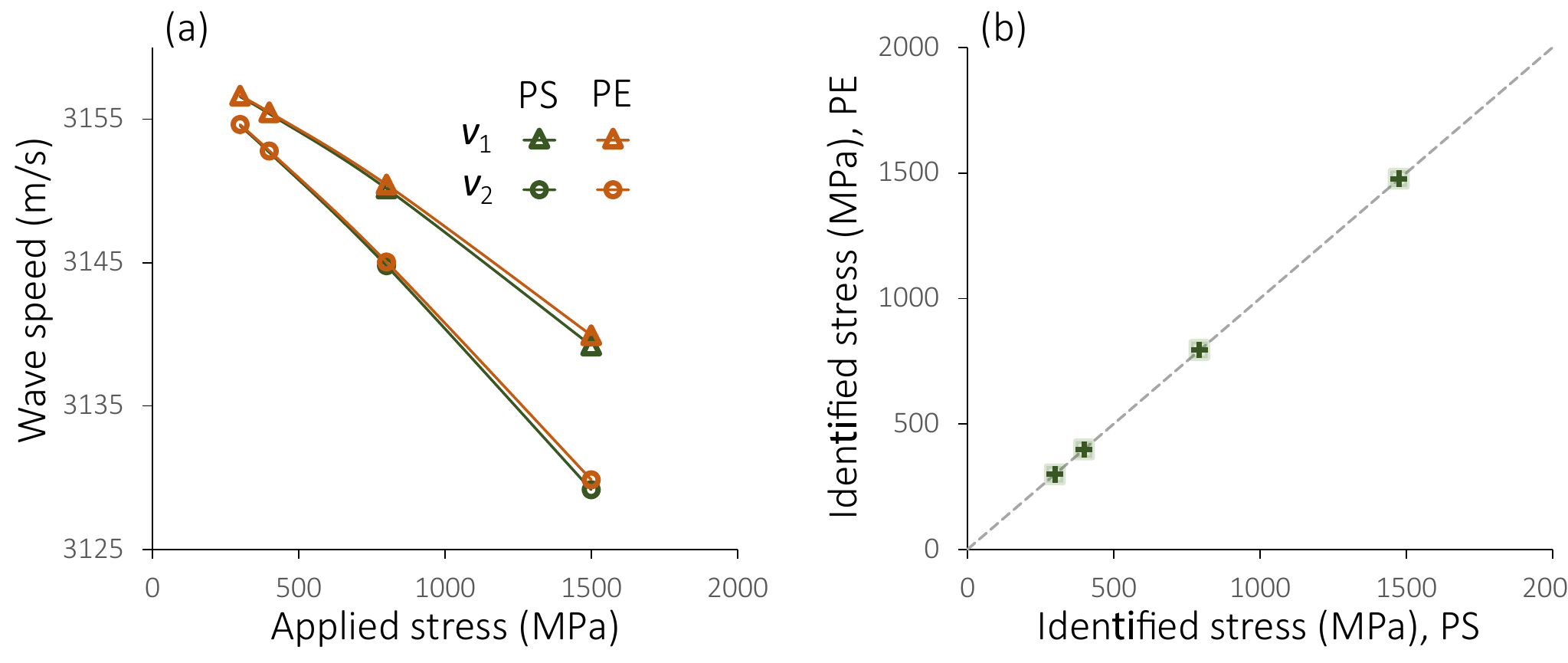}
  \caption{Comparison between plane stress (PS) and plane strain (PE) simulations when the waves travel along $\theta_1=35^\circ$ and $\theta_2=55^\circ$. (a) The speeds $v_1$ and $v_2$ vs the applied stress $\sigma_1$. (b) Comparison of the identified stress obtained from the plane stress and plane strain simulations.}
  \label{fig:PS and PE}
\end{figure}

To achieve PS, we adapt the stretch ratio $\lambda_3$ so we can maintain $\sigma_3 = 0$ when applying a uniaxial stress. To achieve PE we fix $\lambda_3=1.0$ and allow $\sigma_3$ to change.
\Cref{fig:PS and PE}(a) shows the stress along the horizontal $\sigma_1$ against the wave speeds $v_1$ and $v_2$. The wave speeds obtained from the PS model are (slightly) smaller than those obtained from the PE model, and this difference is significant when it comes to predicting the stress.
However, when using these wave speeds to predict the stress with \eqref{eqn:stress-identity-second-order} (see \Cref{fig:PS and PE}(b)), we find that both the PE and PS models predict the same levels of stress.
This indicates that the identity~\eqref{eqn:stress-identity-second-order} is in practice independent from a small amount of out-of-plane stress, which we then confirmed analytically. This independence is highly convenient, because in practice the deformation state may neither be plane stress nor plane strain.

\section{Discussion}

One significant drawback of most ultrasonic methods to measure stress is that they need significant calibration for each specimen. This requires either prior knowledge on the material constants, or complicated measurement systems.

In this paper we propose a solution to these issues by introducing a method that can be used to measure in-plane stress without any calibration. The method requires two angled shear wave speed measurements. At its heart is a formula which is exact for incompressible solids, and applies to compressible solids within the context of initially isotropic third-order elasticity.

The main advantage of our proposed method is that it does not require prior knowledge on the material constants. One challenge in using this method includes fabricating wedges with a precisely chosen inclination: an error of 1\% in the inclination typically results in a error of 4\% when predicting the stress, see~\Cref{fig:angle-errors}. Other aspects that still need investigation include how errors in measuring the wave speeds result in errors in predicting the stress, though this will depend on the specific experimental realisation. Another issue that needs to be considered is how texture anisotropy affects the predictionsof this method~\cite{thompson1986angular}.

We validated the method against Finite Element simulations and found its precision to be excellent, and the predictions to be robust in terms of out-of-plane stress. It remains to be demonstrated if this new method is able to accurately predict stresses in the lab and on the ground.

\section*{Acknowledgements}
Artur Gower is grateful to Prof. Rob Dwyer-Joyce and Prof. Roger Lewis for introducing this problem, and for their collaboration. Artur Gower and Michel Destrade are grateful for partial support from the UKAN grant EPSRC (grant no. EP/R005001/1).


\appendix
\section{The instantaneous elastic moduli}
\label{app:instantaneous-moduli}

The theory and practice of \emph{acousto-elasticity} has a long and distinguished history.
Classic works on the topic include contributions by Brillouin \cite{brillouin1946tenseurs}, Hugues and Kelly \cite{hughes1953second}, Toupin and Bernstein \cite{toupin1961sound}, Hayes and Rivlin \cite{hayes1961surface}, or Biot \cite{biot1965mechanics}.
Later, Ogden \cite{ogden1997non} formalised its equations in a compact and elegant manner within the framework of exact non-linear elasticity.
Hence they obtained the following expressions for the instantaneous elastic moduli in \eqref{eqn:eigensystem}
\begin{align}
& \mathcal A_{01111} = \lambda_1 \frac{\partial \sigma_1}{\partial \lambda_1},
&& \mathcal A_{02222} = \lambda_2 \frac{\partial \sigma_2}{\partial \lambda_2},
&& \mathcal A_{01122} =  \lambda_2 \frac{\partial \sigma_1}{\partial \lambda_1} + \sigma_1,
\notag \\
& \mathcal A_{01212} = \dfrac{\sigma_1 - \sigma_2}{\lambda_1^2 - \lambda_2^2}\lambda_1^2,
&& \mathcal A_{02121} = \dfrac{\sigma_1 - \sigma_2}{\lambda_1^2 - \lambda_2^2}\lambda_2^2,
&& \mathcal A_{01221} = \dfrac{\sigma_1 \lambda_2^2 - \sigma_2 \lambda_1^2}{\lambda_1^2 - \lambda_2^2}.
\end{align}

Now to make progress, we must specify the strain energy density $W$ so that we can compute the Cauchy stress components $\sigma_i = \lambda_i \partial W/\partial \lambda_i$ ($i=1,2,3$, no sum) and their derivatives with respect to the stretch ratios.

As we are focusing on steel, which deforms very little elastically, it is sufficient to use the general strain energy  of third-order weakly nonlinear elasticity.
We take it in its Murnaghan \cite{murnaghan1951finite} form,
\begin{equation} \label{eqn:W-murnaghan}
W = \tfrac{1}{2}(\lambda+ 2\mu)I_1^2 - 2\mu I_2 + \tfrac{1}{3}(\ell + 2m)I_1^3 -2 m I_1 I_2 + nI_3,
\end{equation}
where $\lambda, \mu$ are the Lam\'e constants, $\ell,m,n$ are the Murnaghan constants, and $I_1,I_2,I_3$ are the first three principal invariants of the Green-Lagrange strain tensor $\vec E$. We  align the coordinate axis so that $\vec E$ is diagonal in the $(x_1,x_2,x_3)$ coordinate system, with components $E_i = (\lambda_i^2-1)/2$ ($i=1,2,3$).

The current mass density $\rho$ which is usually not known, is related to  $\rho_0$, the initial mass density (measured before applying stress) through \cite{hughes1953second}:
\begin{equation} \label{eqn:current-density}
  \rho = \rho_0/(1+2I_1+4I_2+8I_3).
\end{equation}

Another commonly used form is the Landau strain energy that uses the (third-order) Landau constants  $A$, $B$, $C$.
They are connected  to the Murnaghan constants $m$, $n$, $\ell$ through
\begin{equation}
\ell = B + C, \qquad m = \frac{A}{2} + B, \qquad n = A.
\end{equation}

Then the stress components and eventually, the instantaneous moduli are straight-forward to compute with a Computer Algebra System.


By assuming small strains, the instantaneous moduli for third-order elastic materials can be written explicitly in terms of the stresses, see  Gower et al. \cite{gower_new_2017} and Tanuma and Man  \cite{tanuma_perturbation_2008}, resulting in \eqref{eqn:A},
where the non-dimensional coefficients $a$, $b$, $c$, $d$ are defined in terms of Murnaghan constants as
\begin{align}
  &a = \frac{2 \ell \mu - \lambda^2 + (n - 2 m - \lambda) (\lambda + \mu)}{\mu (3 \lambda +  2 \mu)},
& & c = \frac{2 \lambda + 2 m - n}{2 \mu},
\\
  & b = - \frac{\lambda n + \mu (4\lambda + 2\mu - 2 m + n )}{2 \mu(3 \lambda + 2  \mu)},
  && d = 2 + \frac{n}{2 \mu}.
\end{align}
For~\eqref{eqn:A} to be asymptotically equivalent to classical third-order elastic models, such as~\eqref{eqn:W-murnaghan}, the constants $b$, $c$, $d$ need to be linked through \cite{gower_new_2017,tanuma_perturbation_2008}:
\begin{equation}
  \lambda (1+ 3b + d) + \mu(1+2b-c) = 0,
\end{equation}
as can be checked easily.

For the material parameters of steel, we use the data collected in Table \ref{tab:material parameters}.
\begin{table}[!ht]
\centering
\begin{tabular}{@{}l|c|c|c|c|c|c@{}}
 & $\rho_0$ & $\lambda$ & $\mu$  & $\ell$ & $m$  & $n$
 \\ \midrule
Rail steel 1~\cite{egle_measurement_1976} & 7800 & 115.8 & 79.9 & -248 & -623 & -714
\\
Rail steel 2~\cite{karim2012measurement} & 7777 & 112.9 & 80.8 & -88.9 & -591.6 & -903.8
\\
Nickel steel~\cite{crecraft1962ultrasonic} &    & 909 & 78.0 & -46 & -590 & -730
\\
Nickel-steel S/NVT~\cite{crecraft1967measurement} &   & 109.0 & 81.7 & -56 & -671 & -785
\\
Hecla 37 carbon steel~\cite{smith1966third} &  7823 & 111 & 82.1 & -461 & -636 & -708
\\
Hecla 17 carbon steel~\cite{smith1966third} &  7825 & 110.5 & 82.0 & -328 & -595 & -668
\\
Hecla 138A steel~\cite{smith1966third} &  7843 & 109 & 81.9 & -426.5 & -619 & -708
\\
REX 535 nickel steel~\cite{smith1966third} &  7065 & 109 & 81.8 & -327.5 & -578 & -676
\end{tabular}
\caption{Material parameters for several samples of steel. The density $\rho_0$ is in kg/m$^3$ and the elastic constants are in  GPa ($\lambda$, $\mu$: second-order Lam\'e constants; $\ell$, $m$, $n$: third-order Murnaghan constants.)}
\label{tab:material parameters}
\end{table}
Whereas \Cref{tab:material parameters-2} gives the  values of the constants $a$, $b$, $c$, $d$ for the Rail steel 1 and Rail steel 2 specimens.
\begin{table}[!ht]
\centering
\begin{tabular}{@{}l|c|c|c|c|c|c@{}}
 & $\lambda $ & $\mu$ & $a$ & $b$  & $c$  & $d$
 \\ \midrule
Rail steel 1~\cite{egle_measurement_1976} & 115.8 & 79.9 & 0.701 & -0.118 & -1.88& -2.47  \\
Rail steel 2~\cite{karim2012measurement} & 112.9 & 80.8 & 0.127 & 0.370 & -0.332 & -3.59
\end{tabular}
\caption{$\lambda$ and $\mu$ are reported in GPa; $a$, $b$, $c$, and $d$ are non-dimensional constants. }
\label{tab:material parameters-2}
\end{table}

\bibliographystyle{unsrt}
\bibliography{reference}

\end{document}